# Poroelasticity of bottlebrush and linear polymer networks

*Nolan A. Miller, Alfred J. Crosby\**

*Department of Polymer Science & Engineering, University of Massachusetts Amherst, Conte Center for Polymer Research, 120 Governors Drive, Amherst, MA 01003, USA*

Email: acrosby@umass.edu

Keywords: polymer network architecture, permeability, solvent transport, relaxation indentation, swollen soft matter

## 1. Abstract

The transport of solvent molecules through soft, swollen networks is critical in both natural and engineered systems. While this poroelastic flow has traditionally been explored in networks where the mesh size is comparable to the solvent molecule size, the effect of network architecture on permeability remains underexplored. Here we investigate solvent transport in linear polymer networks (LPN) and highly elastic bottlebrush elastomer networks (BBN), where the presence of densely grafted sidechains allows for control over swelling and mechanical properties. By synthesizing BBNs with systematically varied crosslinking density ($n_x$) while maintaining constant sidechain length ($n_{sc}$) and grafting density ($n_g$), we probe the poroelastic response in the stretched backbone regime (SBB). Poroelastic relaxation indentation experiments, performed in toluene, reveal how permeability scales with crosslink density and polymer volume fraction. Compared to LPN with identical chemistry, the BBN exhibited a lower permeability scaling exponent with polymer volume fraction that closely matches the theoretical exponent. Despite architectural differences, permeability data for both networks collapse onto a single curve when plotted against dry shear modulus. Our findings demonstrate that molecular network architecture significantly influences permeability, offering new routes to tailor solvent transport in soft, swollen networks. These insights highlight BBNs as a promising platform for applications in permeable membranes, filtration, and microfluidic systems, and pave the way for further studies on how network parameters, such as sidechain length, impact permeability in these highly tunable materials.

## 2. Introduction

Flow is a universal mechanism for transport found in nature and technology at multiple length scales. The transport of nutrients through subsurface streams in an ecosystem[1] are akin to the permeation of oxygen and minerals through blood vessels.[2] With mathematical foundations stemming from understanding soil consolidation and the transport of water,[3,4] poroelastic flow relates the elastic properties and diffusive timescales of a solvent permeating through a porous material. Poroelasticity is used to characterize diffusive properties at the molecular length scale in soft, swollen matter, including synthetic polymer films, brain tissue, and cartilage, using stress-relaxation indentation.[5–9] The description of solvent transport in these systems is built on the assumption that the network size is on a similar order of magnitude to the solvent molecule size, establishing a diffusive flow regime over convective flow. While many studies have focused on the role of polymer chemistry or network size,[10–14] the influence of network architecture stemming from densely grafted systems on solvent transport has not been explored to the best of our knowledge. In a fixed polymer volume fraction, the network architecture modifies the spatial arrangement of polymer, establishing a design space for managing transport and mechanical properties. Here, we investigate how the size of the network, using crosslink density to control the swollen polymer volume fraction, impacts poroelastic permeability using highly elastic bottlebrush elastomer networks (BBN).

The BBN architecture offers advantageous control over the mechanical and swelling properties.[15] The length of the sidechain ($n_{sc}$), grafting density along the backbone ($n_g$), and distance between crosslinks ($n_x$), can be independently tuned to customize the stress-strain response and failure stress of the network.[16] Previous work has used this network parameter triplet, {$n_{sc}$, $n_g$, $n_x$}, and the dimensions of the molecular structures to define the conformation and confinement of the network strands.[17–19] We focus on the stretched backbone (SBB) regime to ensure a system with minimal entanglements.[15,20] BBNs are capable of much larger equilibrium swelling conditions than linear network counterparts with the mitigation of entanglements.[13,21] This capacity to swell is described by Flory-Rehner theory where the osmotic swelling free energy is balanced with the elasticity of the crosslinked network.[22–24] The interplay of elasticity, swelling, porosity, solvent size and shape, viscosity, and crosslinking has been studied in soft materials,[25–32] but few studies have investigated the influence of network architecture on solvent transport.

Poroelastic relaxation indentation is applied here to extract poroelastic constants and calculate permeability as a function of network architecture. The BBNs are compared to poly(dimethylsiloxane) (PDMS, Sylgard 184) linear polymer networks (LPN) to contrast the effects of architecture with that of a commonly used materials system of the same general chemical composition (Figure 1).

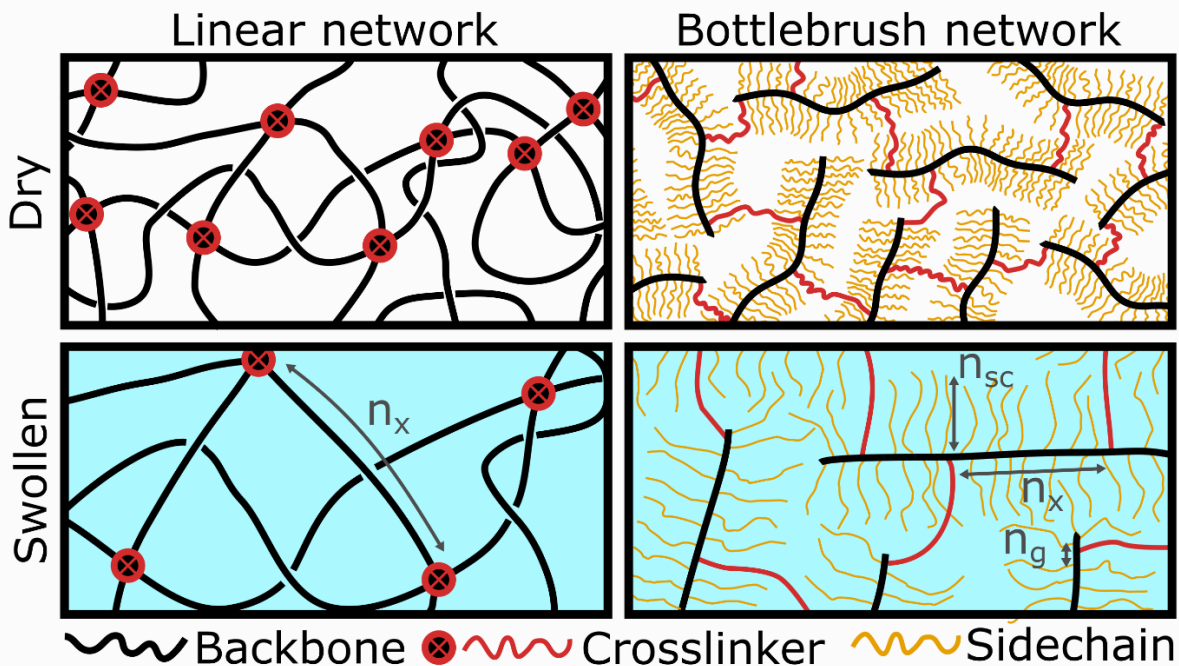

*Figure 1:* Comparison of the linear polymer network (LPN) (left column) and bottlebrush polymer network (BBN) (right column) architectures in a dry (top row) and swollen (bottom row) equilibrium. The sidechains in the BBN increases the effective entanglement molecular weight, resulting in a greater swelling capacity with reduced entanglements.

As the crosslinking density increases, the mesh size in both networks decreases; additionally, the sidechains in the BBNs further impede solvent transport. With control and ease of customization for BBNs, characterizing permeability at the molecular length scale within these highly tunable networks has useful applications for permeable membranes, filtration, and microfluidics.

## 3. Methods

We synthesize solvent-free BBN network architectures with varied $n_x$ while maintaining $n_{sc}$ and $n_g$ constant so that the network remains in the SBB regime.[15] The sidechain is a monohydride-terminated poly(dimethylsiloxane) (Molecular weight (MW) = 4,750 g/mol, MCR-H21, $n_{sc}$ = 60), which was mixed with the crosslinking chain, a hydride-terminated poly(dimethylsiloxane) (MW = 17,200 g/mol, DMS-H25), to control sidechain spacing along the backbone of $n_g$ = 4. The concentration of crosslinking chains was varied to synthesize the four networks with different distance between crosslinkers, {$n_x$ = 60, 30, 12, 6}. The backbone polymer is a trimethylsiloxy terminated, (vinylmethylsiloxane)-dimethylsiloxane copolymer, (MW = 50,000 g/mol, VDT-5035). All mixed networks were poured into a 10 mm thick mold and cured at 70 °C for 48 hours.

Cylindrical slabs with a 10 mm radius were cut from the LPN and BBN of cured material and fully swollen to equilibrium in toluene for at least three days before characterization. Prepared samples were stored in a toluene solvent bath between experiments to prevent fracture stress during drying. The polymer volume fraction, $\varphi_p$, in each swollen network is measured using the mass of the network before and after swelling, shown in equation 1, where ρ and m are density and mass, respectively, and subscripts p and s are polymer and solvent, respectively.

$$\varphi_p = \left(1 + \frac{\rho_p}{\rho_s}\left(\frac{m_s}{m_d} - 1\right)\right) \tag{1}$$

The BBN and LPN samples are labelled as "LPN#" and "BBN#" where # corresponds to the theoretical mesh size of the network (rounded to the nearest nanometer). The swollen elastomer networks were characterized using a compression tester (TA.XTplusC from Stable Microsystems).

The swollen BBN networks were submerged in a solvent bath and a stainless steel, spherical probe with radius, a=12.7 mm, was indented into the sample to an indentation depth of {h=0.6 mm, 0.75 mm, 0.9 mm} (Figure 2a). Each LPN was indented three times to h=0.75 mm. At the start of indentation, the probe is intentionally left out of contact with the surface of the swollen network to measure the capillary force from the solvent contact with the probe and to help identify the initiation of contact for the softest swollen networks. Upon reaching the maximum indentation depth, the poroelastic force relaxation is measured over time as solvent diffuses out from the compressed network.

4. Results

The force-indentation curve, F(h), is fit using fitting parameter, A, to equation (2), a modified form of Hertzian contact to account for a displacement, C, between the surface of the gel and the probe at the start of the experiment.

$$F(h) = A(h - C)^{\frac{3}{2}} \tag{2}$$

The equilibrium swollen shear modulus, $G_s$, is calculated from the fitting parameter, A, and the contact radius, a, using equation (3).[5] The measurements of swollen shear modulus for each crosslink density of the LPNs and BBNs is plotted in Figure 3a & 3b. The network is assumed to be incompressible when swollen to equilibrium where the osmotic free energy from solvent swelling equals the elastic free energy of the network following Flory-Rehner theory.

As the crosslinking density increases, the equilibrium swelling decreases, resulting in a stiffer swollen shear modulus and greater peak force at a given indentation depth. This effect is illustrated using representative curves from each of the bottlebrush networks (Figure 2b). Upon reaching the specified indentation depth, the force response decreases over time as the compressive pressure induced by the probe drives solvent out of the network. The solvent migration from the compressed, swollen network can be characterized by poroelastic properties. The poroelastic relaxation experiment begins once the probe has reached the experimental indentation depth. The force response over time, F(t), is fit using a custom MATLAB script to equation 4 using the initial force at the start of poroelastic relaxation, F(0), the plateau force at long relaxation times, F(∞), and the diffusivity, D, as fitting parameters. The contact radius between the probe and the swollen gel is constant and calculated using the probe radius and experimental indentation depth, accounting for the gap, C, between the probe and the sample at the start of the experiment.

$$F(t) = \bigl(F(0) - F(\infty)\bigr)\left[0.491 \exp\left(-0.908\sqrt{\frac{Dt}{a^2}}\right) + 0.509 \exp\left(-1.679\frac{Dt}{a^2}\right)\right] + F(\infty) \quad (4)$$

The first and last 10% of time from the poroelastic indentation relaxation data was masked for the fitting to remove the effects of viscoelastic relaxation and the force response from withdrawing the probe.

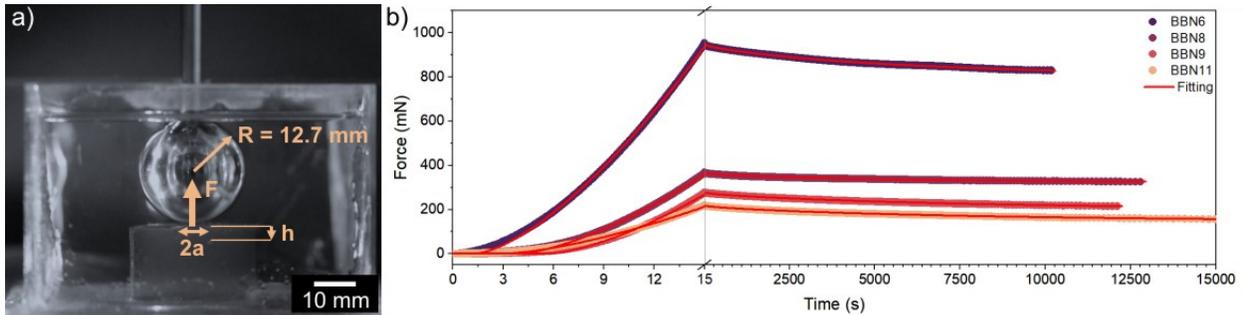

*Figure 2: (a) Representative image of the experimental poroelastic relaxation indentation setup and (b) typical force response over time from indentation and relaxation*

The fit values for F(0) and F(∞) are used to calculate the Poisson's ratio, $\upsilon$, of the drained network using equation 5.[6] The Poisson's ratio for each crosslink density of the LPNs and BBNs is plotted in Figure 3c & 3d.

$$\upsilon = 1 - \frac{F(0)}{2F(\infty)} \quad (5)$$

Finally, the diffusivity is extracted from the fit using equation 4 for each crosslink density of the LPNs and BBNs and is plotted in Figure 3e & 3f. The average diffusion coefficients for the LPN and BBN architectures are (1.23 ± 0.53) × $10^{-9}$ m$^2$/s and (0.73 ± 0.06) × $10^{-9}$ m$^2$/s, respectively. The diffusion coefficient in the BBNs was found to be more consistent than in LPNs.

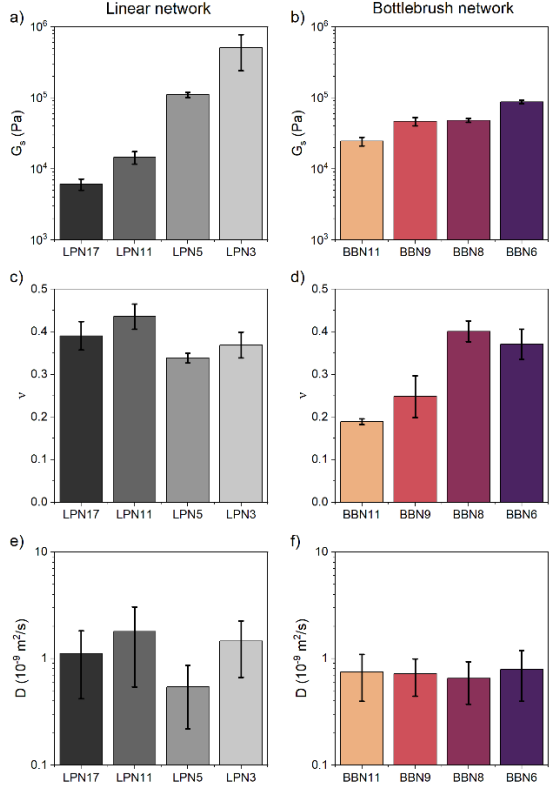

*Figure 3: Poroelastic properties compared between a linear and bottlebrush network architecture with varying degrees of crosslinking that contrast the (a,b) dry shear modulus, (c,d) Poisson's ratio, and (e,f) diffusivity.*

The swollen shear modulus, Poisson's ratio, diffusivity, and solvent viscosity, ($\eta$), define the permeability, (k), of the network using equation 6. We assume the dynamic viscosity of toluene to be 0.56 mPa s.[33]

$$k = \frac{D\eta(1-2\nu)}{2G_s(1-\nu)} \quad (6)$$

Permeability has units of area that represent an effective pore size for solvent transport. We plot the calculated values of permeability for both LPNs and BBNs in Figure 4a as a function of polymer volume fraction and in Figure 4b as a function of swollen shear modulus. As the crosslink density increases, the polymer volume fraction at swollen equilibrium increases. The Kozeny-Carman relationship, shown in equation 7, theorizes that the relationship between the

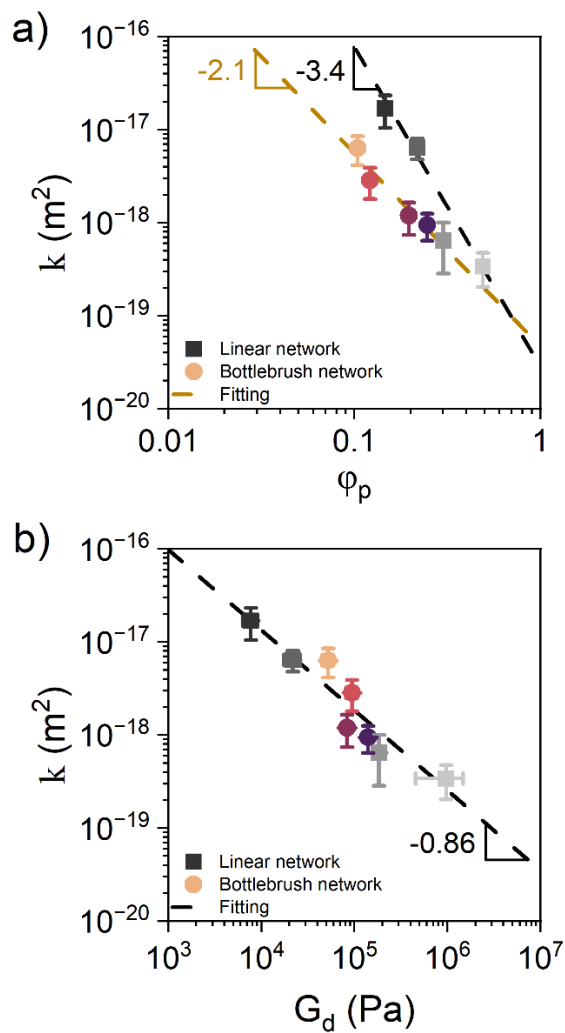

*Figure 4:* Calculated permeability as a function of (a) polymer volume fraction and (b) dry shear modulus. The scaling relationships are independently fit for permeability and the polymer volume fraction. A consolidated fit was used to determine the scaling relationship for permeability and the dry shear modulus.

permeability and the polymer volume fraction is defined by an inverse power-law scaling with exponent, α.[28]

$$k \sim \varphi_p^{-\alpha} \tag{7}$$

The fit exponent for the LPN and BBN is 3.4 ± 0.6 and 2.1 ± 0.4, respectively. The BBNs architecture exhibited an increased resistance to solvent permeability than the LPNs when comparing networks of similar polymer volume fraction. Interestingly, the scaling between permeability and polymer volume fraction for LPNs and BBNs collapse to a single scaling relationship when the permeability of the network is plotted as a function of the dry shear modulus (Figure 4b) with a consolidated fit of 0.86 ± 0.13. These results identify how architecture can be used to independently control the elastic and transport properties of a polymer network.

## 5. Discussion

The concepts within poroelastic relaxation under a constant load were presented by Terzaghi[4] and later generalized by Biot.[3] The framework presented by Biot assumes a linear stress-strain response under small strain limits such that $k \sim G_d^{-1}$, where $G_d$ is the modulus of the drained network. Beyond Biot's and Terzaghi's frameworks, later theories have linked permeability to polymer network parameters through scaling relationships. The Kozeny-Carman theory predicts that $\alpha = 2$ for a porous network while recent literature has shown it can range from 1.4 to 3.3,[34] yet this exponent has not been widely reported for other polymer systems. A recent study measured a Kozeny-Carman exponent of $\alpha = 1.8$ for polyacrylamide gels with $\varphi_p < 0.1$.[28] For our system, the fit exponent for the BBN closely matches predicted Kozeny-Carman exponent. The change in this exponent is understood to be a product of the arrangement of polymer chains in the swollen network, which coincides well with our measured exponent for the LPN where solvent molecules are not impeded by sidechains.

BBN are often cited for having swelling capacities orders of magnitude larger than LPN counterparts in addition to strain-stiffening effects at the larger swelling strains.[25,35–37] The differences between LPN and BBN in the scaling of permeability with polymer volume fraction suggest that polymer architecture offers unique opportunities for controlling solvent transport independent of elasticity. By changing the architecture, from linear networks to densely grafted networks, while maintaining the same volume fraction of polymer, the permeability of the softest networks is reduced by nearly an order of magnitude. Similarly, the permeability of a network could be maintained by changing the network architecture from an LPN to BBN, while using fewer polymer molecules per volume. This capability is set by the mechanism of solvent transport in the two different architectures, as indicated by the consolidated trend of permeability with the dry shear modulus for both LPN and BBN. The changes in the entanglements in both architectures sets these differences, which has also been discussed in scaling theories.[38–41] The ability to have polymer systems with identical chemistry and polymer volume fraction that transport solvent at significantly different rates opens opportunities for engineering systems. For example, adhesion, dependent upon chain density and chemistry, may be provided along with enhanced transport. Our results also suggest that future studies should investigate scaling relationships between permeability and other architecture parameters. For example, increasing $n_{sc}$, within the SBB regime, may further decrease permeability due to the increased impedance for transport.

## 6. Conclusions

We explored the influence of network architecture on the permeability of swollen elastomeric networks, focusing on differences between LPN and BBN as a function of crosslink density. Using poroelastic relaxation indentation, our measurements of permeability revealed a unique interplay between network structure and solvent transport at the molecular level. Comparisons with LPNs highlight how the presence of sidechains in BBNs, which reduces entanglement density and enhances swelling capacity, modulates solvent transport without changing the underlying polymer chemistry. These findings emphasize the potential of BBNs for applications requiring controlled solvent transport, such as membranes, filtration systems, and microfluidic devices.

## 7. Conflicts of Interest

There are no conflicts to declare

## 8. Acknowledgements

This research was funded the U. S. Army Research Laboratory under contract/grant number W911NF-23-2-0022.